\RequirePackage{fix-cm}
\documentclass[twocolumn,epjc3]{svjour3}  
\smartqed  
\RequirePackage{graphicx}
\journalname{Eur. Phys. J. C}
\begin{document}

\title{Spectroscopic study of light scattering in linear alkylbenzene for liquid scintillator neutrino detectors
}


\author{Xiang Zhou\thanksref{e1,addr1}
        \and
        Qian Liu\thanksref{addr2} 
        \and        
        Junbo Han\thanksref{addr3}
        \and
        Zhenyu Zhang\thanksref{addr1}
        \and
        Xuan Zhang\thanksref{addr4}
        \and
        Yayun Ding\thanksref{addr4}
        \and
        Yangheng Zheng\thanksref{addr2}
        \and
        Li Zhou\thanksref{addr4}
        \and
        Jun Cao\thanksref{addr4}
        \and
        Yifang Wang\thanksref{addr4}
}

\thankstext{e1}{e-mail: xiangzhou@whu.edu.cn}


\institute{Hubei Nuclear Solid Physics Key Laboratory, Key Laboratory of Artificial Micro- and Nano-structures of Ministry of Education, and School of Physics and Technology, Wuhan University, Wuhan 430072, China \label{addr1}
           \and
           School of Physics, University of Chinese Academy of Sciences, Beijing 100049, China \label{addr2}
           \and
           Wuhan National High Magnetic Field Center, Huazhong University of Science and Technology, Wuhan 430074, China \label{addr3}
           \and
           Institute of High Energy Physics, Chinese Academy of Sciences, Beijing 100049, China \label{addr4}
}

\date{Received: date / Accepted: date}

\maketitle

\begin{abstract}
We has set up a light scattering spectrometer to study the depolarization of light scattering in linear alkylbenzene. From the scattering spectra it can be unambiguously shown that the depolarized part of light scattering belongs to Rayleigh scattering. The additional depolarized Rayleigh scattering can make the effective transparency of linear alkylbenzene much better than it was expected. Therefore sufficient scintillation photons can transmit through the large liquid scintillator detector of JUNO. Our study is crucial to achieving the unprecedented energy resolution 3\%/$\sqrt{E\mathrm{(MeV)}}$ for JUNO experiment to determine the neutrino mass hierarchy. The spectroscopic method can also be used to judge the attribution of the depolarization of other organic solvents used in neutrino experiments. 
\end{abstract}

\section{Introduction}
\label{intro}
Organic scintillators have played key roles in the 60-year-long research history of neutrinos\cite{Joo}. They had been chosen as the targets in the liquid scintillator detectors where anti-electron neutrinos were firstly discovered by Cowan and Reines\cite{Cowan}, the large mixing angle (LMA) solution of the solar neutrino problem was confirmed by KamLAND experiment\cite{KamLAND}, the neutrino oscillation mixing angle $\theta_{13}$ was precisely measured in Daya Bay experiment\cite{An}, and the spectrum of proton-proton neutrinos was recently observed by Borexino experiment\cite{Borexino}. Because of the unexpectedly large value of $\theta_{13}$, the neutrino mass hierarchy and the leptonic \textit{CP}-violating phase could be measured in the next generation of neutrino oscillation experiments\cite{Cahn}. 

Recently, Jiangmen Underground Neutrino Observatory (JUNO) experiment has been approved to measure the neutrino mass hierarchy by a organic liquid scintillator detector\cite{Li2013}. There will be 20 kiloton liquid scintillator filled in the huge spherical central detector of JUNO whose diameter will be about 35 m. A ternary hydrogenous organic liquid would be chosen as the detector target in JUNO where the solvent is linear alkylbenzene (LAB) whose formula is C$_6$H$_5$C$_n$H$_{2n+1}$ (n=10-13) and the primary and secondary solutes are 2,5-diphenyloxazole (PPO) and 1,4-bis[2-methylstyryl]benzene (bis-MSB), respectively. The reactor anti-neutrino can be detected via the inverse beta-decay reaction $\bar{\nu}_e+p\to e^+ +n$. The positrons and neutrons will product scintillation in LAB. The solutes, PPO and bis-MSB, will transfer the wavelengths of scintillation to longer ones for avoiding the self-absorption of LAB\cite{Xiao}. In the discussion of light transmission in the liquid scintillator, the reference wavelength is usually chosen as 430 nm around which the intensity of scintillation spectrum is maximum. 

JUNO experiment will determine the neutrino mass hierarchy by measuring a very precise and high-statistics reactor anti-neutrino energy spectrum\cite{Li2014}. The method demands an unprecedented energy resolution 3\%/$\sqrt{E\mathrm{(MeV)}}$, i.e., about 1200 photon electrons per MeV\cite{Li2013}. The large scale of the central detector will make scintillation photons transmit several tens of meters before they finally arrive at the photonmultipliers and then transfer into photon electrons. The photons attenuate exponentially when they transmit in a liquid. The detection challenge requires that the solvent of liquid scintillator, LAB, should be high transparent to scintillation photons\cite{Cahn}. After purifications LAB can has a large attenuation length which is about 20 m at 430 nm\cite{Gao}. Even if the scintillation photons vanish exponentially after transmit 20 m, the energy resolution could still hardly reach the requirement of JUNO experiment. Attenuation contains two sub-processes, scattering and absorption\cite{Agostinelli}. Since the liquid scintillator should be dust-free, the scattering process of LAB should be Rayleigh scattering. Because Rayleigh scattering only changes the directions of photons, the scattered photons could still have a change to be detected in a 4$\pi$ detector. It has been found that Rayleigh scattering can increase the effective transparency of scintillation in the large liquid scintillator detector\cite{Suekane}. Simulations have shown that if the Rayleigh scattering length of LAB at 430 nm would be 30 m, the energy resolution of the central detector of JUNO could satisfy the requirement for measurement of neutrino mass hierarchy\cite{Wang}. 

Recently light scattering in several organic solvent including LAB has been studied experimentally\cite{Wurm}. It has been found that light scattering in LAB is depolarized at 90$^\circ$. It is well known that the scattered light at 90$^\circ$ will be completely polarized for Rayleigh scattering of small isotropic dielectric spheres\cite{Rayleigh}. The Rayleigh scattering length of LAB would about 40 m at 430 nm if only considering the polarized part of light scattering\cite{Wurm}. The depolarized part of light scattering in LAB had been treated as a result of absorption/reemission. A significant difference between Rayleigh scattering and absorption/reemission is that the wavelengths of the deflected photons will remain in the former process and change in the latter one. Recently, it has been argued that the depolarized part of light scattering in LAB should also belong to Rayleigh scattering because of the effect of molecular anisotropy. Thus the total scattering length of Rayleigh scattering will be less than 30 m at 430 nm\cite{Zhou}. 

To judge whether the depolarized part of light scattering in LAB belongs to Rayleigh scattering or absorption/reemission is crucial to achieving the unprecedented energy resolution 3\%/$\sqrt{E\mathrm{(MeV)}}$ for JUNO experiment to measure the neutrino mass hierarchy. In this paper, we studied spectroscopically the depolarization of light scattering in LAB. The experimental setup used for the scattering spectrum measurements is described in Sect. \ref{sec2}. Results and discussions of the scattering spectra and the depolarization ratio for LAB are presented in Sect. \ref{sec3}. A summary is given in Sect. \ref{sec4}.

\section{Experimental setup}
\label{sec2}
\begin{figure}[!htp]
\includegraphics[width=8cm]{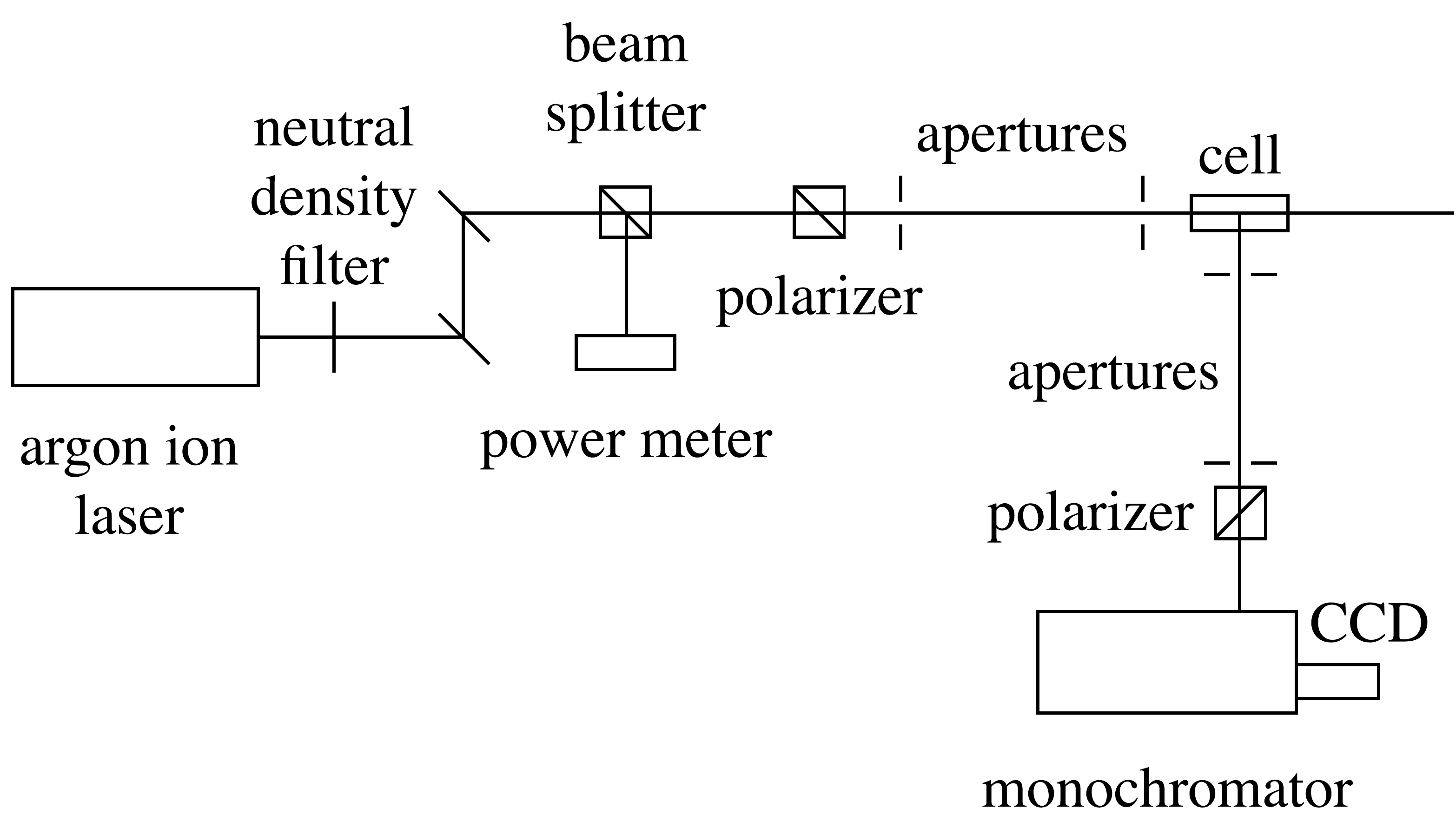}%
\caption{The schematic diagram of the light scattering spectrometer.\label{fig1}}%
\end{figure}

The schematic diagram of the light scattering spectrometer is shown in Fig. \ref{fig1}. The light source was a commercial Spectra-Physics BeamLok 2060 argon ion continuous-wave laser. The beam from the laser had a divergence of 0.45 mrad with a beam diameter of 1.7 mm. The light beam supplied by laser was vertical polarized where the extinction ratio was better then 100:1.

The beam firstly passed a circular continuously variable reflective neutral density filter which can adjust light intensity continuously. The beam was then collimated by two reflection mirrors and two apertures. Between the mirrors and apertures, a beam splitter was used to split the beam into two parts. One part of the beam remained the original path and the other perpendicular one was received by a power meter to monitor the stability of laser beam. Since the optical components would induce spurious depolarization, a Glan Laser calcite polarizing prism was put behind splitter to improve the extinction ratio to be better than 5000:1. After two apertures the laser beam enter the cell whose light path was 50 mm. The cell was of strain-free fused quartz of very high optical quality. The cell was mounted on an optical bench whose height can be adjusted for final alignment of the cell with respected to the incident beam. This alignment was not changed during the experiments. The cell emptied or filled with LAB was held firmly in position during measurements by a removable spring yoke. The attenuation length of the same batch of the LAB sample is about 20 m\cite{Gao}. 

Two apertures whose diameters of holes were 1 mm normal to the laser beam had served to limit the field of view of the spectroscopy and prevent stray light from entering the spectroscopy. The first aperture was 65 mm behind the cell. The second aperture was 790 mm after the first aperture. A Glan-Thompson calcite polarizer was mounted on a motorized precision rotation mount (Thorlabs PRM1/MZ8) which was 115 mm behind the second aperture. The scattered photons entered the slit of the Shamrock SR-500i grating monochromator which was 53 mm behind the calcite polarizer. The light finally collected by the Andor Newton EMCCD DU971P CCD detector. The temperature of the CCD detector had been cooled to -75 $^\circ$C to decrease the noises. The experiment was held in a cleaning room with room temperature controlled at 21$\pm$1$^\circ$C. The scattered photons were collected horizontally and vertically separately by the Glan-Thompson calcite polarizer. The horizontal and vertical scattering spectra can be given by the spectroscopy with the CCD detector.

\section{Results and discussion\label{sec3}}
\begin{figure}[!htp]
\includegraphics[width=9cm]{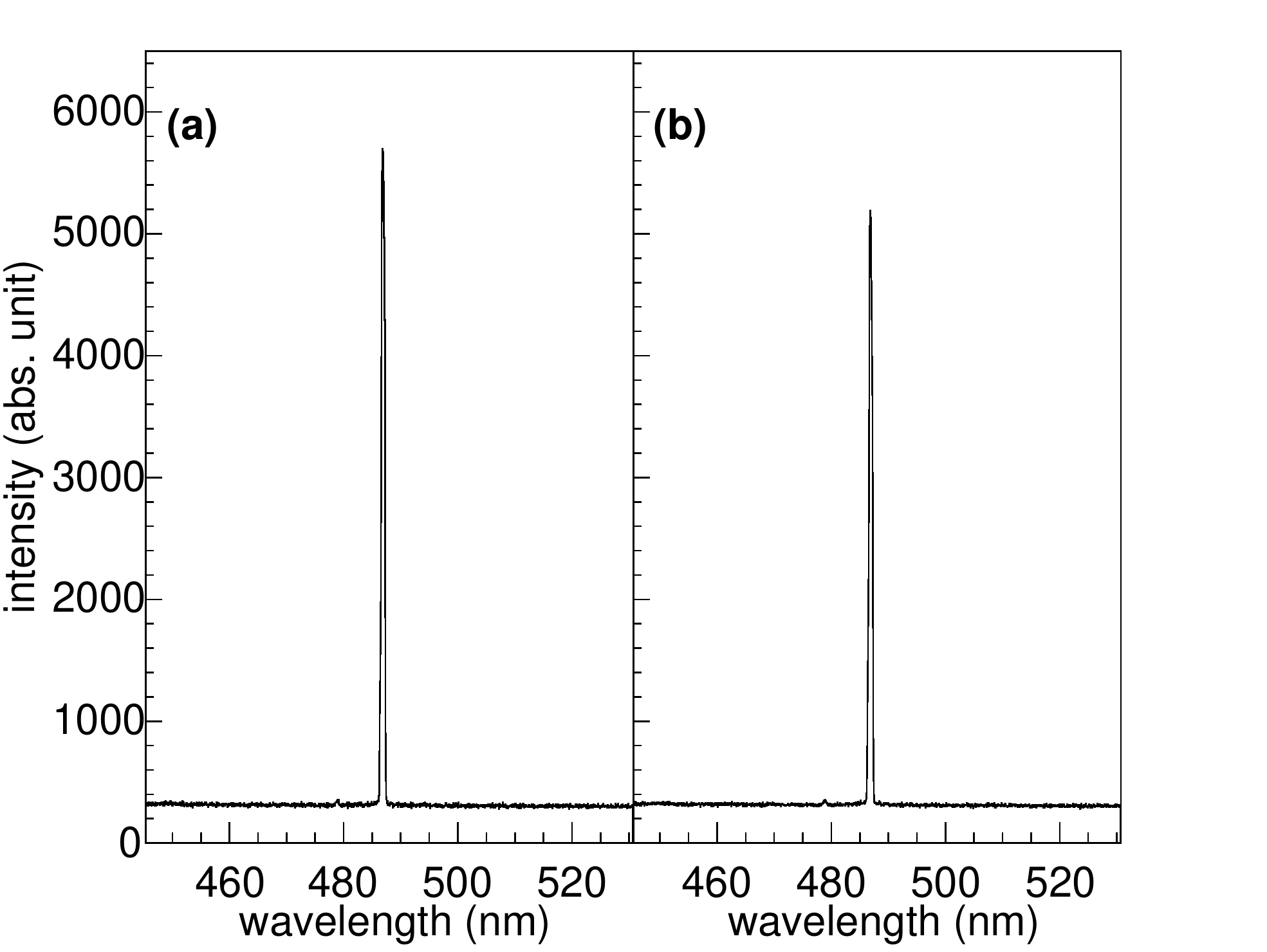}\\
\caption{(a) The vertical scattering spectrum and (b) the horizontal scattering spectrum of the empty cell for 488 nm incident laser beam. The slit width was 0.2 mm and the integral time of CCD was 5 s.\label{fig2}}
\end{figure}

Firstly, the output wavelength of the argon ion laser was set to 488 nm and the cell was empty. The output power was approximately 300 mW at 488 nm. The slit width was set to 0.2 mm and then the spectroscopic resolution would be about 1 nm. The vertical and horizontal scattering spectra for the empty cell with the integral time of 5 s are shown in Fig. \ref{fig2}. Because the neutral filter was reflective and the power of laser at 488 mn was very large, the reflected light was bright and full of the lab. Although we had done a carefully light shield, there were still peaks around 488 nm in the spectra caused by stray light. The maximum intensities of both spectra were near to each other because the stray light had been reflected so many times that it would be almost unpolarized. The flat intensities of scattering spectra away from 488 nm were caused by the CDD detector noise. 

\begin{figure}[!hbp]
\includegraphics[width=9cm]{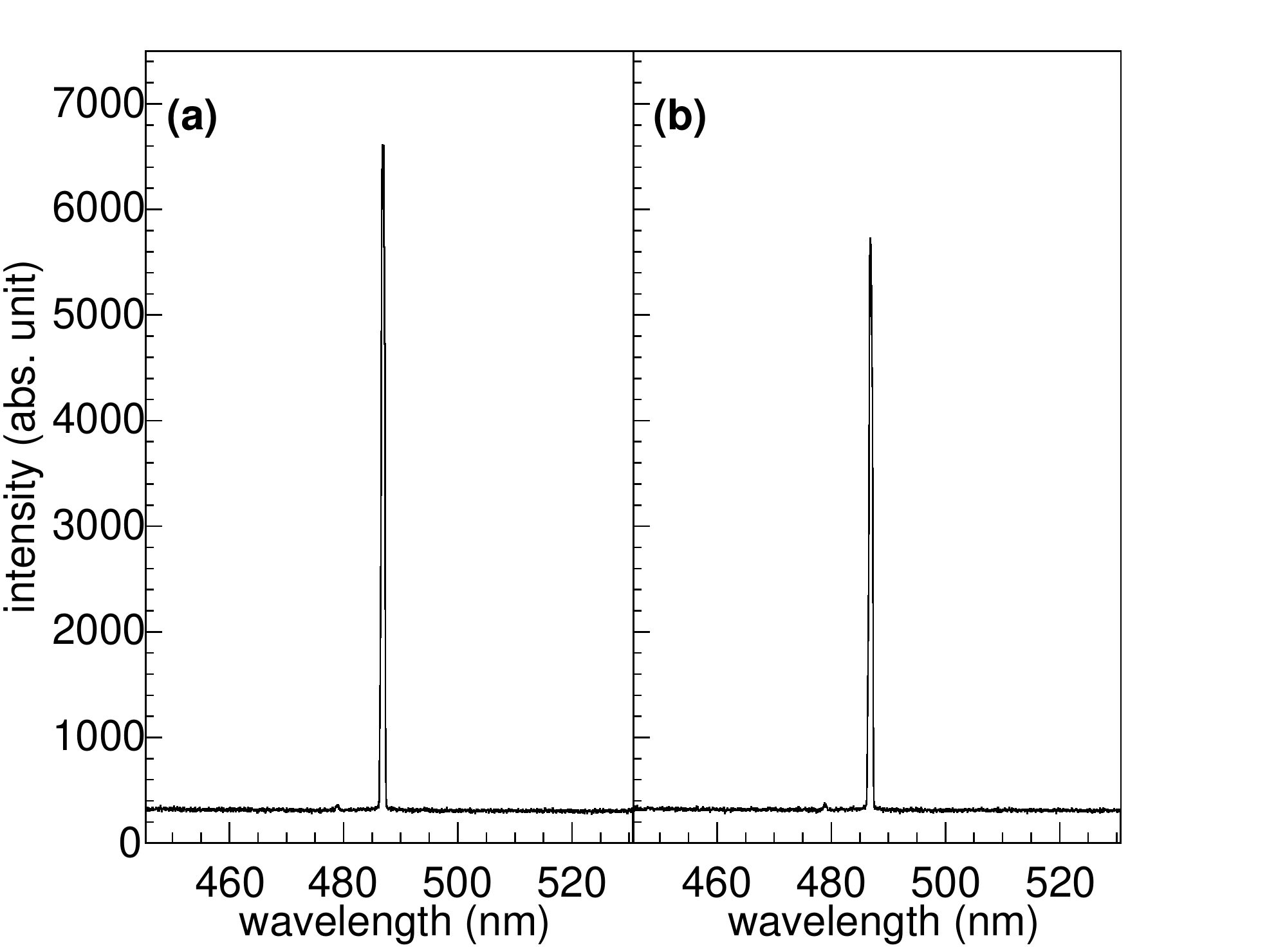}\\%
\caption{(a) The vertical scattering spectrum and (b) the horizontal scattering spectrum of LAB for 488 nm incident laser beam. The slit width was 0.2 mm and the integral time of CCD was 5 s.\label{fig3}}%
\end{figure}

The cell was then filled with LAB without any movement. The vertical and horizontal scattering spectra for LAB with the integral time of 5 s are shown in Fig. \ref{fig3}. Because of light scattering from LAB, the intensities of peaks were larger than those for empty cell. All the wavelengths of peaks in the scattering spectra in Fig. \ref{fig2} and Fig. \ref{fig3} were exactly the same. Thus it can be concluded that the scattered photons from LAB are caused by Rayleigh scattering. 

\begin{figure}[!htp]
\includegraphics[width=9cm]{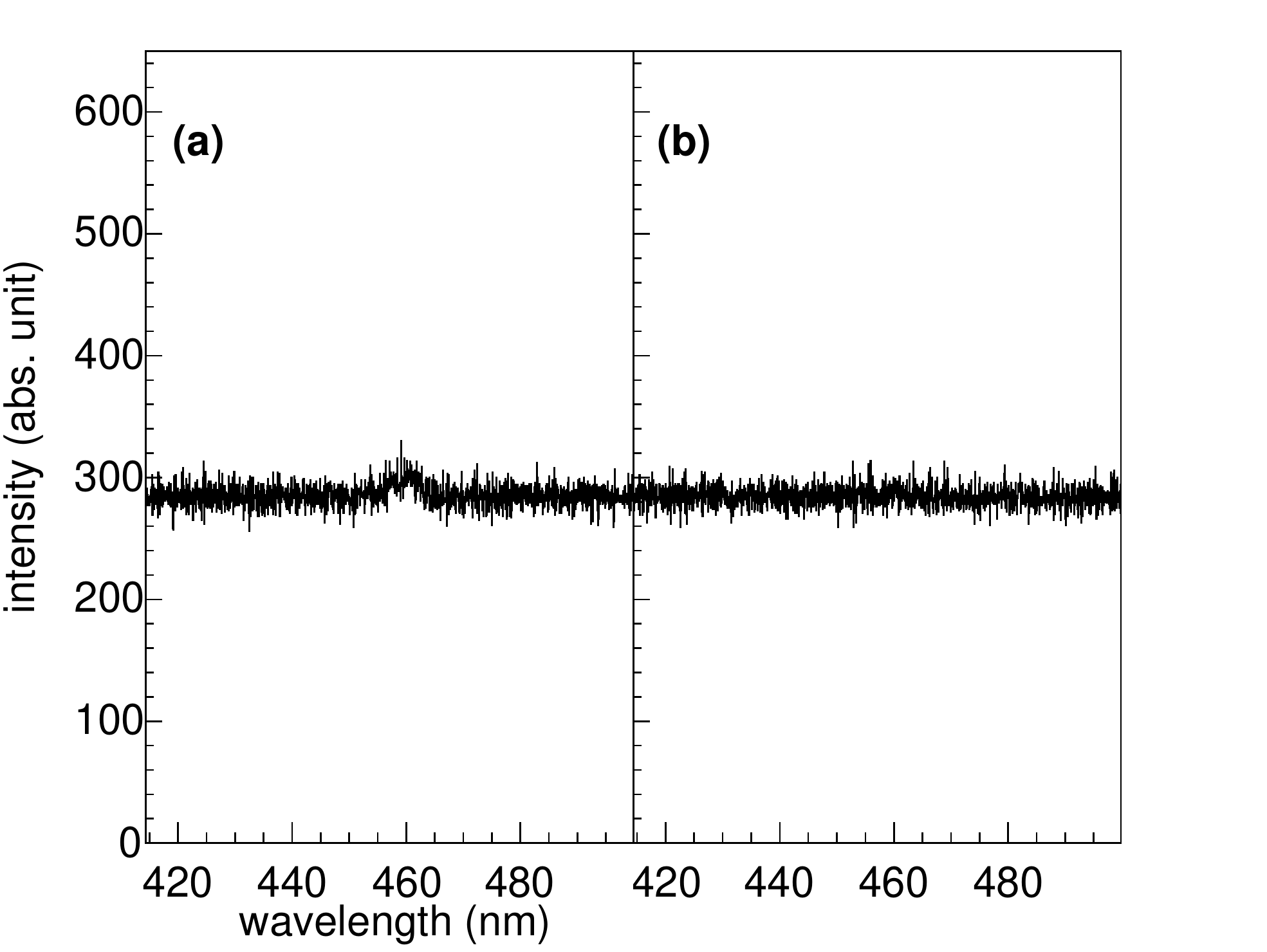}\\%
\caption{(a) The vertical scattering spectrum and (b) the horizontal scattering spectrum of the empty cell for 457.9 nm incident laser beam. The slit width was 1 mm and the integral time of CCD was 20 s.\label{fig4}}
\end{figure}

In the light of the above spectroscopic study, the peak background in spectra caused by stray light has to be suppressed for the quantitative analysis of the depolarization of Rayleigh scattering in LAB. The output wavelength of the argon ion laser was then set to 457.9 nm so that the output power could be lowered to approximately 50 mW. Stray light in lab was successfully suppressed. Meanwhile very few scattered photons could be collected by the spectroscopy for the 0.2 mm narrow slit. The slit width was then widened to 1 mm and the integral time of CCD increased to 20 s. However, the widen slit will lower the spectroscopic resolution to about 5 nm. The vertical and horizontal scattering spectra of the empty cell are shown in Fig. \ref{fig4}. The horizontal scattering spectrum was almost flat, while there was only a very small peak around 457.9 nm in the vertical scattering spectrum which might be caused by light scattering from air or cell. 

\begin{figure}[!htp]
\includegraphics[width=9cm]{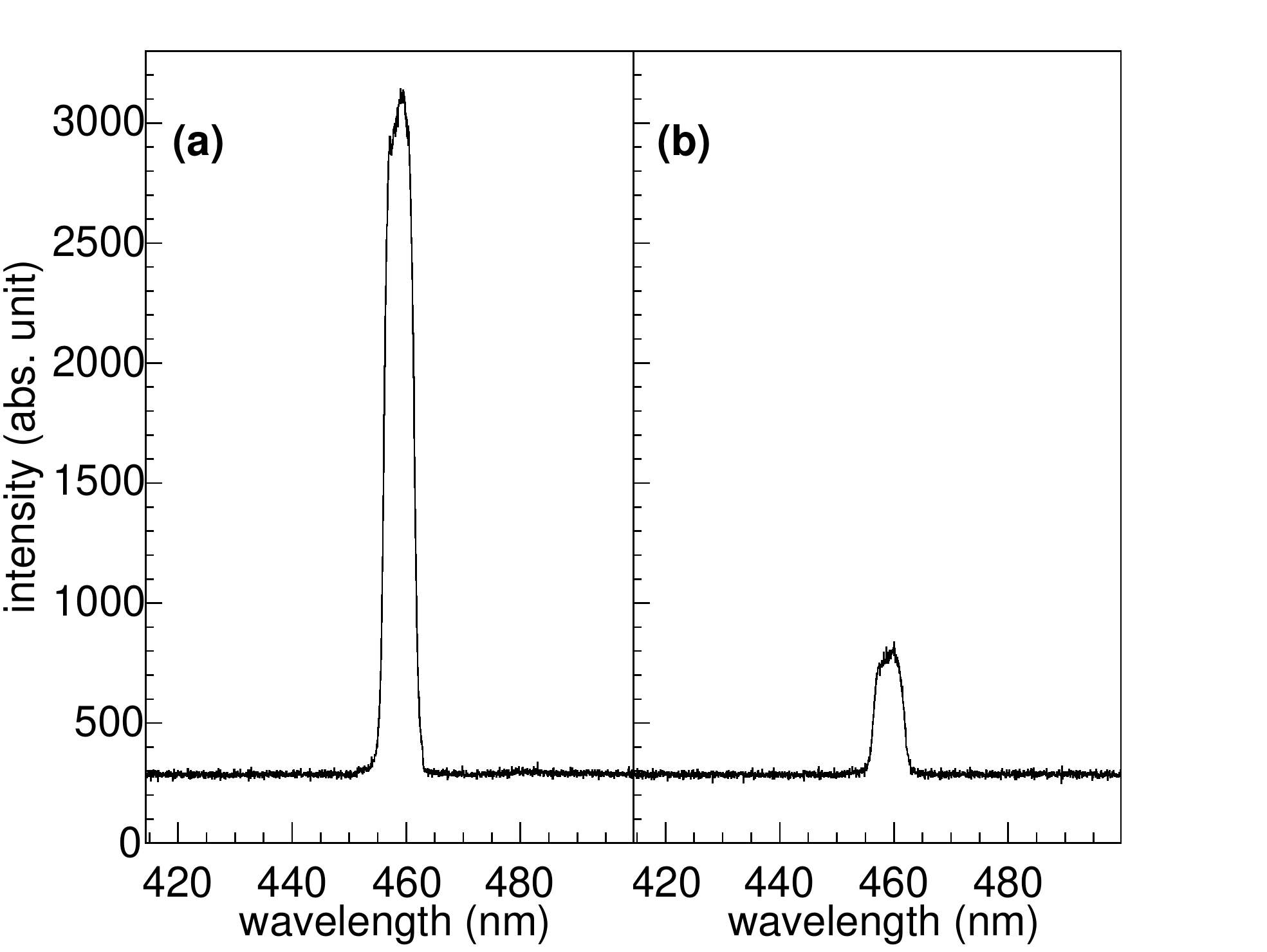}\\%
\caption{(a) The vertical scattering spectrum and (b) the horizontal scattering spectrum of LAB for 457.9 nm incident laser beam. The slit width was 1 mm and the integral time of CCD was 20 s.\label{fig5}}%
\end{figure}

The vertical and horizontal scattering spectra for LAB are shown in Fig. \ref{fig5}. There were obvious peaks around 457.9 nm in the vertical and horizontal scattering spectra. The depolarization of Rayleigh scattering in LAB can be characterized by the depolarization ratio\cite{Kerker}. The directly measured quantity from the scattering spectra is
\begin{equation}
\rho_v=\frac{H_v}{V_v},
\label{eq2}
\end{equation}
where $H_v$ and $V_v$ are the horizontal and vertical intensities of the scattered photons from a vertically polarized incident beam. The depolarization ratio $\rho_u$ of an unpolarized incident beam is defined as
\begin{equation}
\rho_u=\frac{H_v+H_h}{V_h+V_v},
\label{eq3}
\end{equation}
where $H_h$ and $V_h$ are the horizontal and vertical intensities of the scattered photons from a horizontal polarized incident beam. When the scattering angle is 90$^\circ$, $H_v=H_h=V_h$ and then the depolarization ratio $\rho_u$ can be obtain by
\begin{equation}
\rho_u=\frac{2\rho_v}{1+\rho_v},
\label{eq4}
\end{equation}
which is the specific result for the Krishnan's relations at the 90$^\circ$ scattering angle\cite{Krishnan}. 

The scattering intensities $H_v$ and $V_v$ can be obtained from the scattering spectra by subtracting the noise background. The results derived from Eq. (\ref{eq2}) and Eq. (\ref{eq3}) are listed in Table \ref{table1}. The scattering spectra of LAB were measured 3 times. The statistical uncertainty of the depolarization ratio is about 3\%. The bidirectional repeatability of the polarization angle controlled by a precision rotation mount is $\pm$0.1$^\circ$ and it causes the error 1\%. The stability of laser beam had been monitored by a power meter during the measurement and the instability of the laser beam causes the error 1\%. 
\begin{table}[!htp]
\caption{\label{table1} Depolarization ratio of LAB}
\begin{tabular}{ccccc}
\hline
Sample & T          & $\rho_v$(457.9 nm) & $\rho_u$(457.9 nm) & $\rho_u$(430 nm) \\
       & $^\circ$C  & present work          & present work          & Ref.~\cite{Wurm} \\
\hline
LAB    & 21.0       & 0.18$\pm$0.01      & 0.30$\pm$0.02      & 0.31$\pm$0.04   \\
\hline
\end{tabular}
\end{table}

From Table \ref{table1}, the depolarization ratios of 430 nm and 457.9 nm for LAB are the same with each other within 1 $\sigma$. It is well known that the depolarization ratio is a week function of wavelength over the visible range\cite{Kerker,Farinato}. Thus we assess that the depolarized part of light scattering in LAB at 430 nm in the previous experiment belongs to Rayleigh scattering. Thus the total lighting scattering length of Rayleigh scattering in LAB should be less than 30 m at 430 nm\cite{Zhou}. Recently the depolarization ratio of LAB at 405 nm has also been measured independently by a different method whose result is 0.31$\pm$0.02\cite{Liu} which supports our assessment further. 

Experimentally, even gases can have very small depolarization ratios due to molecular anisotropy\cite{Watson}. The light scattered by liquids is generally depolarized\cite{Coumou,Wahid}. It has been used for the polarized Rayleigh scattering to approximately describe light scattering in the water Cherenkov detector\cite{Agostinelli,Shiozawa} and the rare-gas liquid detector\cite{Seidel} because the depolarization ratio of water and rare gas liquids are small. However, the depolarization of Rayleigh scattering in LAB is too large to be ignored. The typical fluorescence time of absorption/reemission for liquid scintillators is several nano-seconds which is sensitive to large detectors\cite{Undaoitia,Lixb}. To treat the depolarized part of Rayleigh scattering as absorption/reemission would increase the time response in simulations of large scintillator detectors which might affect the spatial resolutions and the detection efficiencies\cite{Elisei,Alimonti}. Except for JUNO experiment\cite{Li2014}, LAB has been used in Daya Bay experiment\cite{AnNIM}, RENO experiment\cite{Ahn} and the upcoming SNO+ experiment\cite{Chen2004,Chen2007}. It could also be used in the planned LENA\cite{WurmLENA} and Hanohano observatories\cite{Learned}. 

The previously scattering experiment had also investigate light scattering in phenylxylylethane (PXE) and pseudocumene (PC)\cite{Wurm}. It has been found that the depolarization of light scattering in PXE and PC are even larger than in LAB. PC is currently used as the solvent in the Borexino\cite{Alimonti,Alimonti2009} and KamLAND experiment\cite{Suekane}. PXE has been investigated as a back-up solution for the Borexino experiment\cite{Back}. The diameters of the liquids scintillator detectors in Borexino and KamLAND are 8.5 m and 13 m, respectively. If the depolarized parts would also belong to Rayleigh scattering, the scattering lengths of PXE and PC would shorten to be about 10 m. The spectroscopic method could be used for the Borexino and KamLAND experiments to investigate that whether the depolarization of light scattering in PXE and PC belong to Rayleigh scattering or absorption/reemission.

\section{Summary\label{sec4}}
A light scattering spectrometer has been set up to investigate the depolarization of light scattering in LAB. It has been unambiguously shown that the depolarized light scattering in LAB belongs to Rayleigh scattering. The additional depolarized Rayleigh scattering can increase the effective transparency of LAB. Our study will help JUNO experiment to achieve the unprecedented energy resolution 3\%/$\sqrt{E\mathrm{(MeV)}}$ to determine the neutrino mass hierarchy measurement. Our result is also useful to other neutrino experiments which use LAB as the solvent of the liquid scintillators detectors. The spectroscopic method can be also used to judge the attribution of the depolarization of PXE and PC for the Borexino and KamLAND experiments.

%

\begin{acknowledgements}
This work has been supported by the Major Program of the National Natural Science Foundation of China (Grant No. 11390381), the Strategic Priority Research Program of the Chinese Academy of Sciences (Grant No. XDA10010500), the 985 project of Wuhan University (Grant No. 202273344).
\end{acknowledgements}



\end{document}